\documentclass{wqcd03}                 

\confname{QCD@Work 2003 - International Workshop on QCD, Conversano, Italy, 
14--18 June 2003}

\title{Recent Spin Physics at HERA}

\author{D.~Ryckbosch\addressmark{a}, on behalf of the HERMES Collaboration}

\address[a]{Department of Subatomic Physics, University of Gent, Belgium}

\begin{document}

\begin{abstract}Recent results from the HERMES experiment at HERA are
described. The large data set from Run-I has yielded new information on
the helicity structure of the nucleon. The data to be taken in Run-II will
deal mainly with the transverse spin structure and with exclusive reactions.
\end{abstract}

\maketitle


\section{Introduction}
The flavour structure of nucleons is described in terms of Parton Distribution
Functions (PDF). The unpolarized PDF $q(x,Q^2)$ gives essentially the
probability to find a parton with momentum fraction $x$ in a nucleon. The
polarized PDF's describe the spin structure of the nucleon. In particular
the longitudinal PDF $\Delta q(x,Q^2)$ gives the helicity distribution of 
partons in the nucleon. Unpolarized PDF's have been measured in great detail
over a wide range in $x$ and $Q^2$. The information on polarized PDF's is
much more sketchy. A reasonably accurate picture of the quark helicity 
structure has emerged in recent years, but other aspects of the spin structure
remain virtually untouched. Among these the lack of experimental information
on the gluon polarization and the transverse spin structure of the 
nucleon are particularly obvious. Several experiments are gearing up to
address these issues in the near future.

This paper reports on some recent results obtained by the HERMES experiment
in connection with the longitudinal spin structure of the nucleon, which was
the main topic for the HERMES Run-I (1995-2000). The plans for the Run-II
(2002-2006+) are discussed in the second half of this contribution. Here
the main interest lies in exclusive reactions using a new recoil detector,
and the transverse spin structure of the nucleon.

\section{The HERMES experiment}
HERMES is a fixed target experiment~\cite{spectro}
 on the 27.6 GeV HERA-$e$ ring. The
electrons in the ring are transversely polarized by the Sokolov-Ternov effect,
and longitudinal polarization is achieved  by a spin rotator in front of
the HERMES target. A second rotator behind the experiment
restores the transverse
polarization. Typical beam polarizations are between 50 and 60\%. The
polarization of the beam is continuously monitored by two independent Compton
backscattering polarimeters.

The beam traverses the HERMES target  consisting of a 40 cm
long windowless storage cell which confines the injected polarized target gas 
to the region around the beam. The polarized atomic H and D are provided by
an atomic beam source using Stern-Gerlach separation.
A small sample of the target gas is continuously analyzed in
a Breit-Rabi polarimeter.  Typical polarizations are  85\%. 
In 1996 and 1997 data were taken on
a polarized proton target, while in 1998, 1999 and 2000 a large data sample
was collected on a polarized deuterium target.

The scattered electrons, and some of 
the hadrons produced in the reaction, are detected
in the HERMES spectrometer. This is a typical forward magnetic spectrometer
consisting of two symmetric halves above and below the plane of the 
accelerator. Scattered leptons and produced hadrons are detected and 
identified within the angular acceptance of $\pm 170$ mrad horizontally
and $40-140$ mrad vertically. The hadron-lepton separation is done on the
basis of the signals in a transition radiation detector, an electromagnetic
calorimeter and a pair of scintillator hodoscopes where the second one is
preceded by a lead sheet preshower.

The identification of pions, kaons and (anti)protons
 is performed since 1998 with a Ring Imaging 
CHerenkov detector (RICH) using clear aerogel and C$_4$F$_{10}$ gas as
radiators~\cite{RICH}. The hadrons can be separated over almost the full 
momentum range of HERMES, i.e. between 2 and 15 GeV.
Before 1998 a threshold \v{C}erenkov was used to give
pion identification in a limited momentum range.

The typical kinematic domain for the analysis of polarized DIS in HERMES
is $0.1<Q^2<15$ GeV$^2$, $x>0.02$, with $Q^2$ the (negative) momentum
transfer squared, and $x=Q^2/2M\nu$ the Bjorken scaling variable. To avoid
regions with too large contributions of resonance production or radiative
corrections further conditions  $W>2$ GeV, $y=\nu/E>0.85$ are usually 
imposed, with
$W^2=2M\nu+M^2-Q^2$ the invariant mass (squared) of the photon-nucleon system
and $y$ the fraction of the initial beam energy $E$ transferred. 
In semi-inclusive reactions where 
in addition to the scattered lepton also a produced hadron is detected
the fraction of the energy of the virtual photon carried by the
hadron is given by $z=E_h/\nu$. To avoid the region with predominantly
target fragmentation a cut at $z>0.2$ is usually used.

\section{Inclusive scattering}
\begin{figure}[h]
\hbox to\hsize{\hss
\includegraphics[width=\hsize]{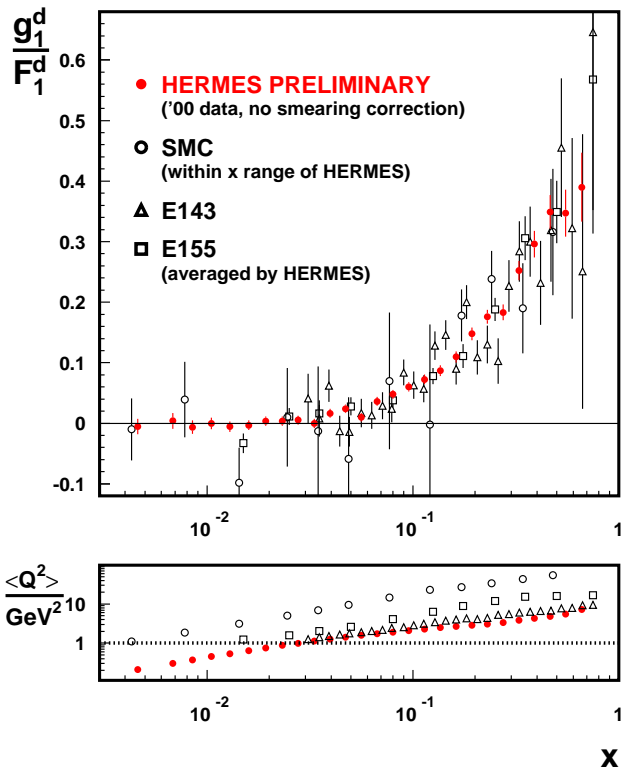}
\hss}
\caption{The ratio of the polarized to the unpolarized structure 
function for deuterium $g_1^d/F_1^d$. The bottom panel shows the $Q^2$
values corresponding to each $x$-bin for the different experiments.}
\label{fig1}
\end{figure}

Most of the information on the PDF's has up to now come from inclusive
DIS: experiments where only the scattered lepton is detected. Fig.~\ref{fig1}
shows the world data on the polarized structure function $g_1(x)$ for
 deuterium. 

It is obvious from this figure that all experiments are basically in agreement
with each other. The recent preliminary HERMES data considerably
 improve the statistical
accuracy. However, there
is a difference of about one order of magnitude in the scale of the various
experiments, with the SMC results being taken at much higher beam energy than
the data from SLAC and HERMES. The fact that the ratio is similar in the
different experiments thus implies an evolution for $g_1$ that is very
similar to the one for $F_1$. Just as was done succesfully in the case of
unpolarized structure functions one can analyse the scaling violations
observed in $g_1(x,Q^2)$ for proton and neutron targets, and deduce 
information on the polarization of the different quark flavours and the
gluon polarization.

\begin{figure}[h]
\hbox to\hsize{\hss
\includegraphics[width=\hsize]{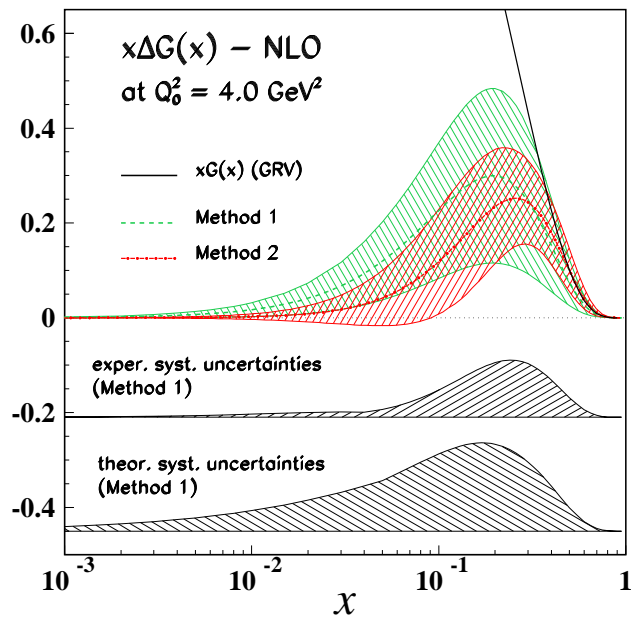}
\hss}
\caption{Gluon polarization as derived from a NLO QCD fit to the world data
on $g_1$ \cite{lara}. ``Method 1'' and ``Method 2'' refer to two different
methods to solve the evolution equations. The full line gives the positivity
limit.}
\label{fig2}
\end{figure}

Such a NLO QCD analysis was recently undertaken by HERMES~\cite{lara}. 
The results confirm the facts already known about the quark polarization:
the quark spin contributes only 20-30\% to the spin of the nucleon; most
of this contribution comes from the valence quarks; the light quark sea is
only little (and negatively) polarized. The gluon polarization that can be
derived from the analysis of the presently available data is shown in
Fig.~\ref{fig2}.

This figure illustrates that the present data do not really constrain the
gluon polarization. The only strong suggestion is that it should be positive,
but no value can be given. More direct methods to determine the gluon
contribution to the nucleon spin are necessary. This is one of the 
aims of the COMPASS experiment at CERN which is now starting up. Also the
spin physics programme at RHIC intends to determine the gluon polarization
independently.

\section{Semi-inclusive scattering}
\subsection{Data and analysis}
The HERMES experiment was specifically designed to perform
 accurate measurements
of semi-inclusive reactions, where apart from the scattered lepton also some
of the produced hadrons are detected. The idea is that these hadrons contain
extra information on the quark that took part in the scattering process.
This technique of flavour tagging allows the determination of the polarization
of individual quark flavours directly from the spin asymmetries observed in
hadron production.

The quantity of interest here is the photon-nucleon asymmetry $A_1^h$ when a 
hadron of type $h$ is produced, which can be derived from the experimental
semi-inclusive spin asymmetry $A_\parallel^h$:
\noindent
\begin{center}
\begin{equation}
A_1^h=\frac{A_\parallel^h}{D(1+\eta\gamma)}=\frac{1}{D(1+\eta\gamma)}
\frac{N^{\uparrow\downarrow}_h-N^{\uparrow\uparrow}_h}{N^{\uparrow\downarrow}_h+N^{\uparrow\uparrow}_h}
\label{a1h}
\end{equation}
\end{center}
where $D$ is the virtual photon depolarization, $\eta$ and $\gamma$ are
kinematic factors, and $N_h^{\uparrow\uparrow} (N_h^{\uparrow\downarrow})$ are
the number of semi-inclusive events (properly taking into account polarization
of target and beam, and relative luminosity) for target polarization 
parallel (antiparallel) to the beam polarization.

\begin{figure}[h]
\hbox to\hsize{\hss
\includegraphics[width=\hsize]{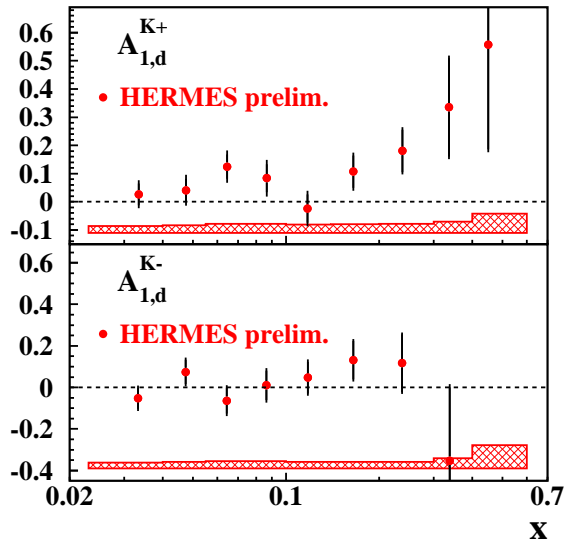}
\hss}
\caption{Semi-inclusive spin asymmetries for kaons produced on a deuteron
target.}
\label{asym}
\end{figure}

An example of the asymmetries obtained at HERMES is given in Fig.~\ref{asym}.
Similar asymmetries were derived for positive and negative pions and on a 
proton target. 
A striking feature of the asymmetries in this figure is the fact that the
asymmetry for the $K^-$ is compatible with zero. Since this is an all-sea
object this already indicates that the polarization of the quark sea
will be very small.

In leading order QCD and assuming factorization of the cross section one can
write this asymmetry in terms of the PDF and fragmentation functions:
\noindent
\begin{center}
\begin{eqnarray}
A_1^h(x,z)&=\frac{\int_{z_m}^1 dz\sum_q e_q^2\Delta q(x)D_q^h(z)}
{\int_{z_m}^1 dz\sum_q e_q^2 q(x)D_q^h(z)}\\
&=\sum_q P_q^h(x)\frac{\Delta q(x)}{q(x)}
\label{a1h}
\end{eqnarray}
\end{center}
In the last equation the {\em purities} $P_q^h$ are introduced. These are
spin-independent quantities which give the probability that a hadron of type
$h$ observed in the final state came from a struck quark of flavour $q$. In 
that sense they are the inverse of fragmentation functions. The purities
depend on the unpolarized quark densities and the fragmentation functions.
The former were measured to high precision in unpolarized DIS experiments.
Information on the fragmentation functions is, however, less precise. There
is some information for pions, but those data are taken at quite different
kinematics than relevant for HERMES. Hence, the purities were calculated
using a Monte Carlo simulation of the entire scattering process. Standard
unpolarized parton distribution parametrizations~\cite{cteq}
 were used, while the
fragmentation was modelled in the LUND string model as implemented in 
JETSET~\cite{jetset}. 
The LUND model fully describes the fragmentation from the current
as well as the target fragmentation region. The parameters of the model
were tuned to fit the hadron multiplicities measured at HERMES in order
to achieve a good description of the fragmentation process at our kinematics.
The resulting purities also include effects of the acceptance of the
spectrometer.

The analysis applies Eq.~\ref{a1h} in matrix form:
\noindent
\begin{center}
\begin{equation}
\vec{A_1}(x)={\cal{P}}(x)\cdot\vec{Q}(x)
\label{matrix}
\end{equation}
\end{center}
where the elements of $\vec{A_1}$ are the measured inclusive and semi-inclusive
(Born) asymmetries, the vector $\vec{Q}$ contains the quark (and anti-quark)
polarizations and the matrix $\cal{P}$ contains as elements the effective
integrated purities for the proton and neutron targets. This is an 
over-constrained system of equations which is solved by a minimisation
procedure.

\subsection{Results}
\begin{figure}[h]
\hbox to\hsize{\hss
\includegraphics[width=\hsize]{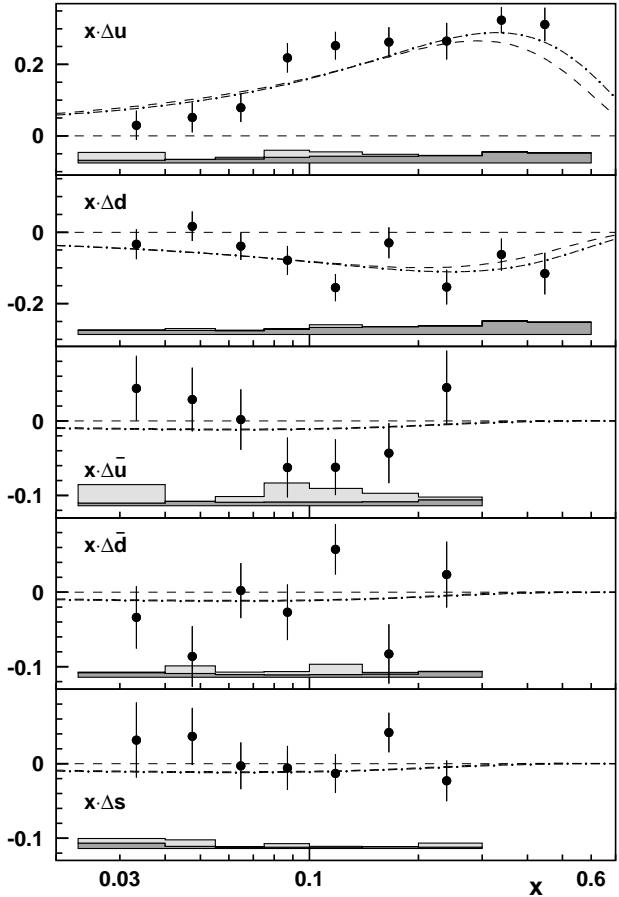}
\hss}
\caption{The $x$-weighted quark spin densities. The dashed line shows a
GRSV parameterization~\cite{grsv}, 
and the dash-dotted curve an alternate parameterization
due to Bl\"{u}mlein and B\"{o}ttcher~\cite{bb}.}
\label{5qdens}
\end{figure}

The purity formalism has been used in the HERMES analysis~\cite{deltaq} to make a flavour 
decomposition of the quark polarizations for $u,\bar{u},d,\bar{d},$ and
$s+\bar{s}$. A symmetric strange sea polarization 
$\Delta s/s=\Delta\bar{s}/\bar{s}$ was assumed. 
The measured asymmetries were integrated over the $z$-range
from 0.2 to 0.8, and over $Q^2>1$GeV$^2$. The resulting spin densities
are shown in Fig.~\ref{5qdens}. 

The polarization of the $u$-quark is positive over the entire measured range
in $x$, with the largest polarization at high $x$ where the valence quarks
dominate. The polarization of the down-quark is negative and also most
important in the region of the valence quarks. The polarization of the
light sea flavours $\bar{u}$ and $\bar{d}$, and the polarization of the
strange sea are consistent with zero.

The data are compared to polarized PDF's derived from fits to inclusive
data. Good agreement is found between the spin densities directly measured
here and the results of NLO QCD analyses, in particular for the valence
distributions. The tendency of the parameterizations to yield a negative sea
polarization is not confirmed by the data, not is it ruled out with the
present statistical accuracy.

\begin{figure}[h]
\hbox to\hsize{\hss
\includegraphics[width=\hsize]{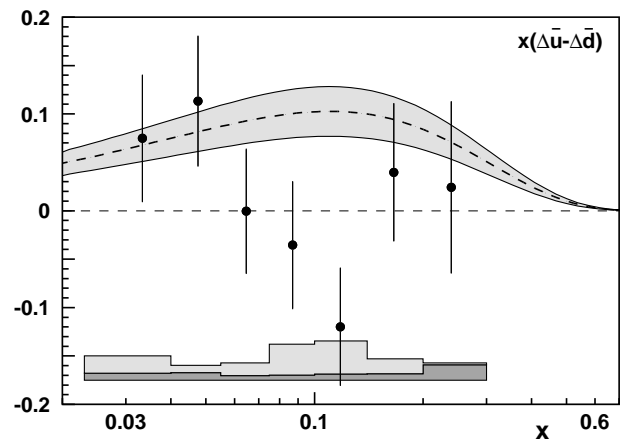}
\hss}
\caption{Flavour asymmetry $\Delta\bar{u}-\Delta\bar{d}$ of the light sea 
extracted from the HERMES purity analysis. The curves show predictions of
the Chiral soliton model~\cite{chsm}.}
\label{du-db}
\end{figure}

The flavour symmetry of the unpolarized light sea is known to be
broken. This is experimentally well established through a violation of the
Gottfried sum rule. Several models which give a good description of this
symmetry breaking in the unpolarized sector also predict a sizeable symmetry
breaking in the polarized sector. In particular, most models 
predict that $\Delta\bar{u}-\Delta\bar{d}>0$. 
The flavour asymmetry 
derived from the present measurement is shown in Fig.~\ref{du-db}.

The present data do not favour a very strong flavour symmetry breaking. 
However, the statistical uncertainties are still rather large and further
measurements of this quantity will be of great interest.

\section{Tensor polarization in deuterium}
When analyzing the experimental inclusive spin asymmetries for deuterium
in order to deduce the structure function $g_1^d$ one has to take into
account a small effect due to possible tensor polarization in this spin 1
target. This is connected with the presence in a spin 1 target of an
additional tensor polarized structure function $b_1$ which relates to the
other structure functions $F_1$ and $g_1$ as:
\noindent
\begin{center}
\begin{eqnarray}
F_1 &=& \frac{1}{3}\sum_qe_q^2(q^++q^-+q^0)\\
g_1 &=& \frac{1}{2}\sum_qe_q^2(q^+-q^-)\\
b_1 &=& \frac{1}{2}\sum_qe_q^2[2q^0-(q^++q^-)]
\label{b1}
\end{eqnarray}
\end{center}
where $q^+$ and $q^-$ denote the probability of finding a quark with helicity
parallel to that of the nucleon and $q^0$ is the probability to find a
quark with momentum fraction $x$ in a nucleon with the target in a 
helicity state 0. $b_1$ thus measures the difference in parton distributions
between a $m=1$ and a $m=0$ target.

\begin{figure}[h]
\hbox to\hsize{\hss
\includegraphics[width=\hsize]{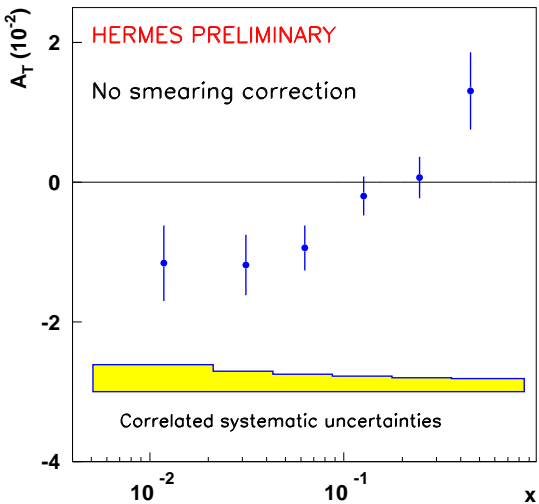}
\hss}
\caption{Tensor asymmetry for deuterium.}
\label{at_fig}
\end{figure}

\begin{figure}[h]
\hbox to\hsize{\hss
\includegraphics[width=\hsize]{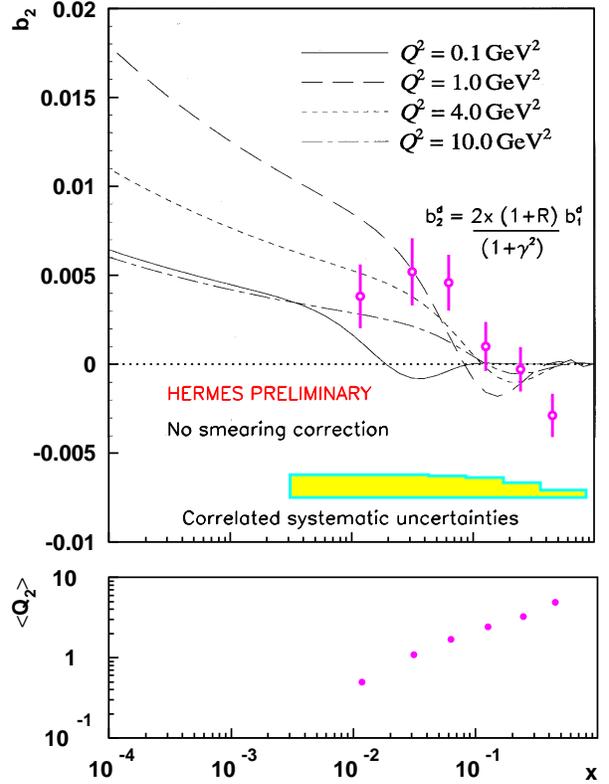}
\hss}
\caption{Tensor structure function $b_2^d$. The lines show the predictions
of Ref.~\cite{bora} for different values of $Q^2$.}
\label{b2}
\end{figure}

The HERMES target is capable of producing highly tensor polarized deuterium.
This was done during a short run in 2000 with an
average polarization of 83\%. The tensor polarized structure function can be
found from the experimental tensor asymmetry:
\noindent
\begin{center}
\begin{equation}
A_T=\frac{(\sigma^++\sigma^-)-2\sigma^0}{\sigma^++\sigma^-+\sigma^0}
\propto -\frac{2}{3}\frac{b_1}{F_1}
\label{at}
\end{equation}
\end{center}
The preliminary HERMES result for this asymmetry is shown in Fig.~\ref{at_fig},
while the deduced tensor structure function $b_2$ (trivially related to
$b_1$ through: $b_2=2x(1+R)b_1/(1+\gamma^2)$) is shown in Fig.~\ref{b2}.
The small values found for the tensor asymmetry (of the order 1\%) show
that the effect on the determination of the helicity structure function $g_1$
can safely be neglected. The actual tensor structure function $b_2^d$ is
significantly different from zero, in particular at lower valus of $x$. It
is in broad agreement with the model calculations of Ref.~\cite{bora}.

\section{Exclusive reactions}
Exclusive DIS  reactions, where the target nucleon remains in or close to its
ground state, can be described in terms of the Generalized Parton Distributions
(GPD) that were introduced a few years ago. These GPD's form a natural
off-forward extension of the standard Parton Distribution Functions
which are well determined from (semi-)inclusive DIS reactions. They form a
connection between the PDF's and the form-factors of hadrons.

\begin{figure}[h]
\hbox to\hsize{\hss
\includegraphics[width=\hsize]{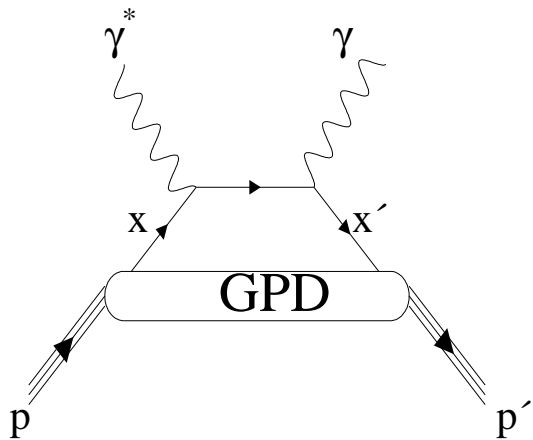}
\hss}
\caption{Handbag diagram describing DVCS.}
\label{handbag}
\end{figure}

The cleanest example of the appearance of GPD's 
in an exclusive reaction is that
of Deeply Virtual Compton Scattering (DVCS).
The main diagram for this reaction is shown in 
Fig.~\ref{handbag}.
There are 4 (flavour sets of) GPD's: 2 unpolarized
functions ($H$ and $E$) and 2 polarized ($\tilde{H}$ and $\tilde{E}$). The
two functions $E$ and $\tilde{E}$ correspond to helicity flip operators and
have no direct analogon in the forward PDF's. The functions $H$ and $\tilde{H}$
have as their forward limit the standard unpolarized and polarized PDF's
$F_1$ and $g_1$, respectively.
The
GPD's depend on 3 independent kinematic variables: $H(x,\xi,t)$ where $\xi$
is related to the usual Bjorken variable by $2\xi=\frac{x_B}{1-x_B/2}$. (Note
that there are other equivalent choices possible for the variables.)

The importance of GPD's in the context of spin physics is embodied by the
sum rule which was derived by Ji~\cite{ji} for the second moment of the GPD's:
\noindent
\begin{center}
\begin{equation}
\frac{1}{2}\int_{-1}^{+1}x[H^q(x,\xi,t=0)+E^q(x,\xi,t=0)]dx = J_q
\label{ji}
\end{equation}
\end{center}
where $J_q$ is the {\em total} angular momentum contribution of quarks to the
spin of the nucleon. Up to now there is experimental
information only on the contribution from the spins of quarks (and gluons).
Combined with this sum rule GPD's and thus exclusive reactions are
 the only way presently known that may
lead to information on the orbital angular momentum contribution, thus
providing a complete picture of the nucleon spin structure.

The main problem in the experimental determination of the GPD's is the fact
that they do not appear as simple factors in the expressions for cross
sections, asymmetries etc. They always appear in convolutions over the
$x$ and $\xi$ variables, making a direct determination impossible. Finding
the best deconvolution procedure to lead from the experimental data to the
GPD is a challenge for the future. To do this a good understanding of the
structure of the GPD is essential to develop adequate models.

\begin{figure}[h]
\hbox to\hsize{\hss
\includegraphics[width=\hsize]{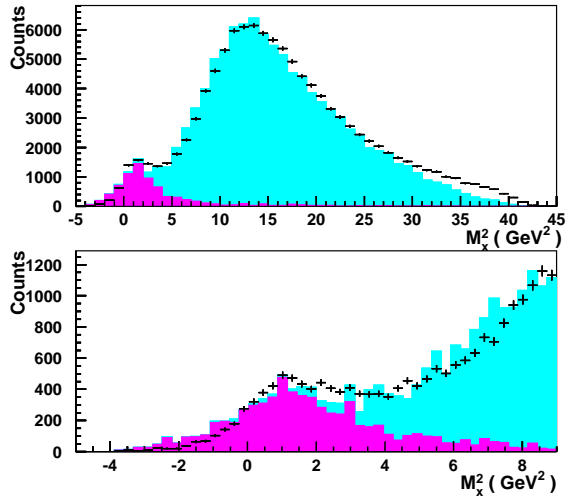}
\hss}
\caption{Missing mass distribution for electroproduction of real 
photons~\cite{olddvcs}. The
dark histogram shows the simulated contribution of exclusive photonproduction,
while the light histogram corresponds to photons produced in fragmentation
processes (mainly $\pi^0$ production).}
\label{dvcs1}
\end{figure}

Fig.~\ref{dvcs1} shows the missing mass (squared) spectrum for the DVCS
reaction at HERMES. A clear peak at the ground state mass can be seen, thus
establishing the observation of exclusive DVCS. 
However, at the HERMES kinematics with relatively low beam energy the main
source of hard photons is not DVCS but the Bethe-Heitler process where the
photon is radiated from either the incoming or the outgoing electron. It
is in principle impossible to disentangle these two processes. On the other
hand the presence of two processes will also lead to interference between
them and this can be exploited to access the DVCS amplitude. The interference
terms can be projected out by measuring azimuthal asymmetries in single
photon electroproduction. It was shown~\cite{dvcs} 
that beam-spin ($A_{LU}$), target-spin
($A_{UL}$) and beam-charge ($A_C$) azimuthal asymmetries provide access to
different combinations of the real and imaginary parts of the interfering
amplitudes. It should be noted that at present HERA is the
only place where beam charge asymmetries can be studied because of the
availability of both electron and positron beams.

\begin{figure}[h]
\hbox to\hsize{\hss
\includegraphics[width=\hsize]{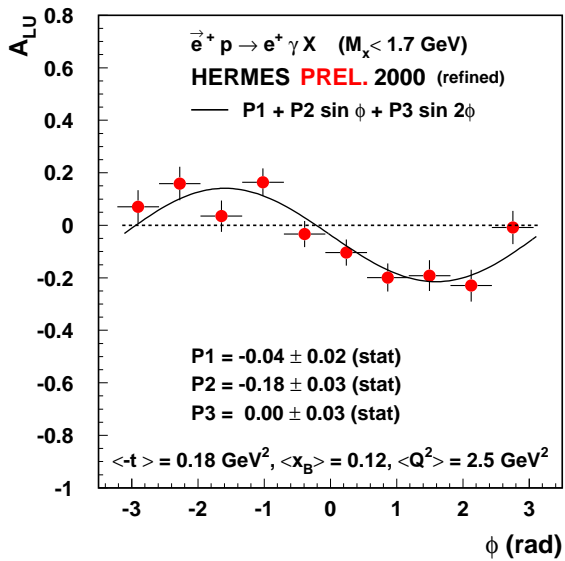}
\hss}
\caption{The beam-spin azimuthal asymmetry for exclusive
real photon production on a hydrogen target.}
\label{dvcs2}
\end{figure}

\begin{figure}[h]
\hbox to\hsize{\hss
\includegraphics[width=\hsize]{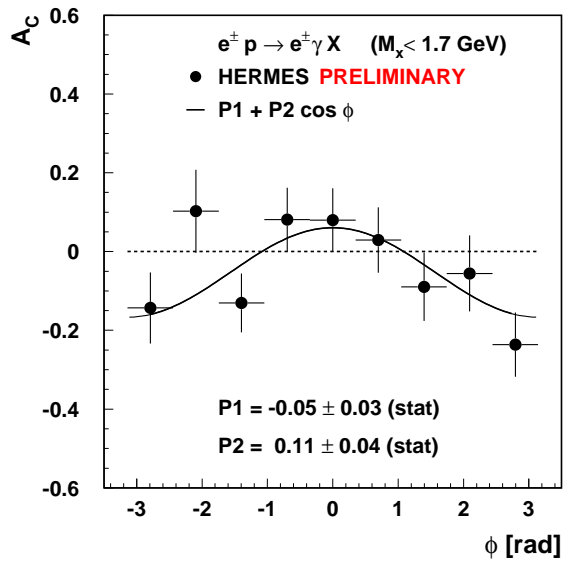}
\hss}
\caption{The beam-charge azimuthal asymmetry for exclusive
real photon production on a hydrogen target.}
\label{dvcs3}
\end{figure}

In Fig.~\ref{dvcs2} the beam-spin azimuthal asymmetry for a proton target
is shown, while Fig.~\ref{dvcs3} presents the results for the beam-charge
asymmetry on the proton. The $\sin\phi$ moment of the former one is related
to the imaginary part of the interference amplitude; the $\cos\phi$ moment
of the latter is related to the real part of the interference amplitude.
It is obvious from the results presented in these figures that the DVCS-BH
interference indeed dominates the distributions. The extracted $\sin\phi$
and $\cos\phi$ moments are in encouraging agreement with theoretical 
calculations based on models of GPD's. In a next step the kinematic
distributions of the various moments of the azimuthal asymmetries are
extracted and to be compared to more detailed calculations.

From Fig.~\ref{dvcs1} it can be seen that the experimental resolution of the
HERMES spectrometer is insufficient to allow a clear separation of exclusive
and fragmentation processes. At present only the scattered electron and the
real photon are detected in the forward spectrometer. To really establish
exclusivity on the event level would require the detection of the recoil
target proton. Kinematics dictate that this proton recoils with low momentum
at high angles relative to the beam line, outside the present acceptance of
the HERMES spectrometer. A major project at HERMES now is the construction
of a recoil detector to detect and identify such recoil particles. The
detector will consist of three active detector parts. A silicon detector
around the target cell inside the beam vacuum, a scintillating fibre tracker
in a longitudinal magnetic field and a layer of scintillator strips with
interspersed W-sheets as preshower material.

It is planned to install the recoil detector somewhere in 2004 or 2005 and
to operate it for on the order of 2 years with an unpolarized hydrogen target
and polarized electron and positron beams. The higher target densities that
are possible for unpolarized targets will enable HERMES to obtain a large
data sample on DVCS with unprecedented statistical and systematic accuracy
for both beam-spin and beam-charge asymmetries. An added advantage of
the recoil detector is that the determination of the momentum of the recoil
proton immediately yields an accurate measurement of the momentum transfer
$-t$ to the target. This will improve the resolution in this important
kinematic quantity (see e.g. Eq.~\ref{ji}) by about an order of magnitude.

\section{Transversity}
Up to now we have only discussed the helicity structure of the nucleon. In
fact there are three different structure functions which describe the
parton structure of the nucleon at leading twist. These are the unpolarized
structure function $F_1$, the helicity structure function $g_1$ and a new 
quantity, transversity $h_1$. The related PDF is denoted $h_1^q(x)$ and
gives the probability to find a transversely polarized quark in a 
transversely polarized nucleon. It is experimentally unknown, mainly because
as a chiral-odd object it cannot be measured through inclusive DIS. In 
combination with another chiral-odd object, e.g. a fragmentation function,
transversity becomes accessible. In semi-inclusive reactions it leads
to single-spin azimuthal asymmetries in e.g. pion and kaon production on
a transversely polarized target.

Such a transversely polarized target is installed in HERMES since 2001.
First data have been taken with it after HERA became operational again 
following a major luminosity upgrade for the collider experiments. However,
at the present time the collected data sample is too small to make a
measurement of the transversity distributions feasible. This will come
in the run of 2003 and 2004. On the other hand, HERMES has already collected a
large amount of data on a longitudinally polarized target. Such a target is
longitudinal with respect to the incoming electron beam, but does contain
a small ($\approx 10-15\%$) transverse component with respect to the
virtual photon. Moreover, there is a number of higher-twist distribution
functions and fragmentation functions which are directly related to the
transversity distribution and which also cause single-spin asymmetries
in semi-inclusive reactions on a longitudinal target.

\begin{figure}[h]
\hbox to\hsize{\hss
\includegraphics[width=\hsize]{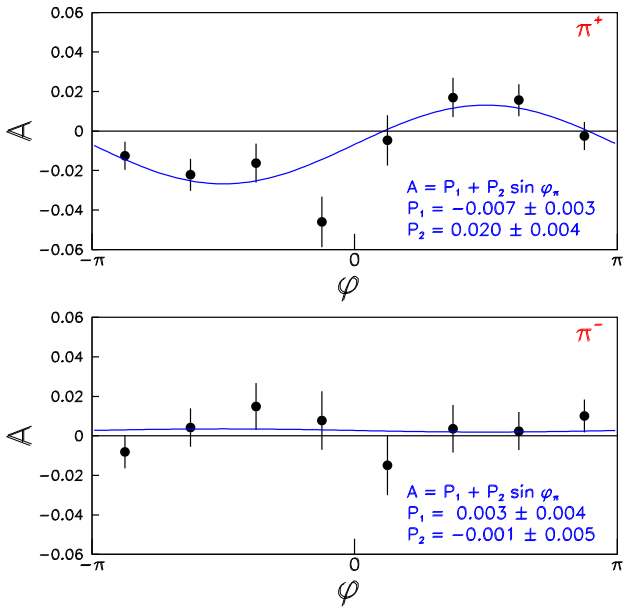}
\hss}
\caption{$\phi$-dependence of the single-spin azimuthal asymmetry in
semi-inclusive pion production on a longitudinally polarized proton target.}
\label{ssa}
\end{figure}

The experimental single-spin azimuthal asymmetry for positive and negative
pions on a longitudinally polarized proton target 
is depicted in Fig.~\ref{ssa}. The positive pions show a clear
$\sin\phi$ modulation of the asymmetry, while this is absent for negative
pions. This is consistent with the picture of $u$-quark dominance.
A larger  set of results on such asymmetries for a
deuteron target was recently published~\cite{ssad}. Theoretical calculations
based on models of the transversity distribution have had some success in
reproducing these data.

\section{Conclusions}
Some recent results from the HERMES experiment at HERA were presented.
The analysis of the large data set on longitudinally polarized proton and
deuteron targets, collected in the first run of HERMES from 1995 through 2000,
is nearing completion. Results of a NLO QCD analysis of the inclusive 
scattering show indications for a positive and possibly large gluon
polarization. The results from the semi-inclusive analysis are consistent
with those from the inclusive data and moreover give for the first time
a flavour decomposition of the quark polarizations in 5 components. The
valence quarks are seen to dominate the contribution of the quark spins
to the nucleon spin, while the sea appears to have a negligible polarization.

The observation of azimuthal asymmetries in exclusive reactions, here
exemplified by the DVCS reaction, promises access to the new GPD's. The
implementation of the recoil detector now under construction at HERMES
will give a much enhanced accuracy for such studies.

The main thrust of the data taking of HERMES at the present time is with a
transversely polarized proton target. On the basis of
 single-spin azimuthal asymmetries
observed in semi-inclusive scattering off a longitudinally polarized
target it can be expected that these data will give first experimental
information on the transversity distribution of quarks in the nucleon.


\end{document}